\documentclass[journal,compsoc]{IEEEtran}
\usepackage{algorithm}
\usepackage[noend]{algpseudocode}

\usepackage{blindtext}
\usepackage{graphicx}
\usepackage{url}
\usepackage{mathtools}
\DeclarePairedDelimiter{\ceil}{\lceil}{\rceil}

\ifCLASSINFOpdf

\else

\fi

\begin{document}

\title{Multi-Level Spherical Locality Sensitive Hashing For Approximate Near Neighbors}

\author{\IEEEauthorblockN{Teresa Nicole Brooks\IEEEauthorrefmark{2} and Rania Almajalid\IEEEauthorrefmark{1}}
\IEEEauthorblockA{
\\Seidenberg School of Computer Science and Information System\\
Pace University\\
New York City, NY\\
Email: \IEEEauthorrefmark{2}tb93141n@pace.edu,
\IEEEauthorrefmark{1}ra56319p@pace.edu
}}

\maketitle

\begin{abstract}
     This paper introduces ``Multi-Level Spherical LSH'': parameter-free, a multi-level, data-dependant Locality Sensitive Hashing data structure for solving the Approximate Near Neighbors Problem (ANN). This data structure uses a modified version of a multi-probe adaptive querying algorithm, with the potential of achieving a $O(n^p + t)$ query run time, for all inputs n where $t <= n$.
\end{abstract}


\section{Introduction}
Locality Sensitive Hashing (LSH) it is a probabilistic, search algorithm that uses hashing to detect similar or nearest neighboring data points using the high probability of hash collisions between indexed data points and a given query. LSH seeks to limit the search space to data points that are highly likely to be similar. Similarity is typically measured as the distance of one data point to another, the closer these points are, the more likely they similar. 

Traditional LSH data structures have several parameters whose optimal values  depend on the distribution distance from a query to a given set of points. The data structure presented in this paper does not rely on such parameters and only adopts the space parameter that the user provides. In this paper we introduce Multi-Level Spherical LSH; a multilevel, parameter-free, data-dependant Locality Sensitive Hashing data structure for solving the Approximate Near Neighbors Problem (ANN). Our data structure is a modified version of the Spherical LSH data structure presented by \cite{Andoni2015a}. To query this data structure we will use the multi-probe adaptive querying algorithm presented by Thomas et al \cite{Andoni2015}. We propose that this data structure could achieve a $O(n^p + t)$ query run time, for all inputs n where $t <= n$. Where the given values t is the number of close points, n is the number of total data points and p is the parameter that describes the "quality of the LSH family used \cite{Andoni2015a}". To our knowledge, the data structure presented in this paper is the first parameter-free, multi-level data-dependant LSH.

\section{Background:}

Locality Sensitive Hashing (LSH) is a probabilistic, search algorithm that uses hashing to detect similar documents via the use of collisions.  One approach to LSH is similar to using minhash, where we seek to hash shingles n times but in this case we want a hashing function that will generate similar hash codes for similar shingles. Using these hashes we can determine which document pairs should be compared for similarity, as only similar documents should have hash collisions. This bucketing of similar items to be compared drastically reduces the search space, hence eliminating the need to examine all document candidate pairs \cite{Rajaraman2011}. 

LSH is used to solve a wide variety of problems such as near duplicate and duplicate document detection. Search engines, news aggregators, e-commerce sites and host of other software applications use algorithms to detect similar documents. Other applications of Locality Sensitive Hashing are nearest neighbor search and spherical range reporting\cite{LSHQBitsBlog}.



\section{Related Works }
This paper explores an open problem presented in "Parameter-free Locality Sensitive Hashing for Spherical Range Reporting" by Thomas D. Ahle, Martin Aumüller, and Rasmus Pagh \cite{Ahle2016} to achieve their desired query run time complexity. In this section, we briefly discuss three key papers that our approach directly builds upon.  These papers are Parameter-free Locality Sensitive Hashing for Spherical Range Reporting\cite{Ahle2016}, Practical and Optimal LSH for Angular Distance \cite{Andoni2015} and Optimal Data-Dependent Hashing for Approximate Near Neighbors \cite{Andoni2015a}.

\subsection{Parameter-free Locality Sensitive Hashing for Spherical Range Reporting}
In this paper, the authors propose two adaptive LSH based algorithms to solve the spherical range reporting problem, as well as a parameter-less LSH data structure. This data structured is a modification of a LSH data structure implemented in the Locality-Sensitive Hashing (LSH) framework of Indyk and Motwani, and it represents the asymptotically best solution to neighboring problems in high dimensional spaces. Traditional LSH data structures have several parameters whose optimal values  depend on the distribution distance from a query to a given set of points. Like the LSH data structure presented by Thomas et al  \cite{Ahle2016}, the data structure proposed in this paper does not rely on such parameters and only adopts the space parameter that the user provides. However, their predicted query time is similar to that of an LSH data structure whose data has been optimally selected for the query and data in consideration, under given space constraints. 

The former best query running time in high dimensional spaces was $\Omega(tn^p )$, which was attained by traditionally LSH-based data structures for outputting a single data point. In their research Thomas et al, implemented a parameter-free LSH data structure which utilizes multi-probing querying. The authors proved that their multi-probe algorithm and parameter free data structure was able to get close to a $O(n^p + t)$ query run time \cite{Ahle2016}. The authors concluded that both their adaptive multi-probe and single-probe algorithms will work at the very least, as well as standard Locality Sensitive Hashing algorithms for a given data set and query for given space restrictions, in relation to the number of repetitions that are set by the user. Because these respective values are never larger than the expected running times, this leads to a modest increase in repetitions per level of the data structure. More specifically, single level data structures require $O(log log n)$ more repetitions and multi-level data structures require $O(log^3 n)$ more repetitions \cite{Ahle2016}.

The theorems presented for both the multi-probe and single probe algorithms proved that the algorithms are never ``more than log factors away from and ideal query time of a tuned LSH data structure'' \cite{Ahle2016}.

Our work and the work of Thomas et al, is based on the prior, highly influential works of A. Andoni, P. Indyk, T. Laarhoven, I. Razenshteyn and L. Schmidt. 

\subsection{Practical and Optimal LSH for Angular Distance}

Practical and Optimal LSH for Angular Distance \cite{Andoni2015} describes and analyzes the underlying algorithms implemented in FALCONN \cite{Razenshteyn}, a library which reflects the approaches first proposed in the Locality Sensitive Hashing Framework of Indyk and Motwani. In this paper they proved there exists a Locality-Sensitive-Hashing (LSH) family for approximated Near Neighbor Search algorithm that yields an optimal running time exponent p for angular distance.  

The primary motivation was to find a class of LSH hash functions that not only have theoretically optimal guarantees like Spherical LSH but are also practical in implementation and application like Hyperplane LSH. Their model which includes a mutli-probe scheme for cross-polytope LSH combined with the techniques of feature hashing and implementing the cross-polytope rotations using Fast Hadamard Transform, indeed produced results that yield theoretical guarantees while making noteworthy, empirically proven improvements over Hyperplane LSH query times for exact nearest neighbor search.

Results from experiments using both random data and real data sets show that their model achieved for data sets with $10^5$ to $10^8$ points using multil-probe query for cross-polytope LSH yielded 10x faster run times for nearest neighbor queries than an optimized implementation of Hyperplane LSH \cite{Andoni2015}.

\subsection{Optimal Data-Dependent Hashing for Approximate Near Neighbors}
In Optimal Data-Dependent Hashing for Approximate Near Neighbors \cite{Andoni2015a} paper, the authors present an optimal data-dependent hashing scheme for the approximate near neighbor problem. To goal, given a set of points in a d-dimensional space 
is to construct a data structure that can retrieve any point within a certain distance to a given query.

This approach is done via data-dependent LSH families, where randomized hashing functions are chosen based on the given data set. The use of such data-dependent LSH families enables Spherical LSH to achieve an optimal p for the ANN problem. In this data structure, data is partitioned into spherical caps using a certain radius and diameter.  This process of data partitioning is usually set-up iteratively and runs a number of times. In each iteration Spherical LSH encloses the data into small balls using a small radius. In this case the smaller the balls are, the better the p value that can be achieved; where $p=log(1/p1)/log(1/p2)$ expresses how sensitive a hash function is to changes in distance. P1 is the collision probability for the near points and p2 is the collision probability for the far points.

Their model achieved $O(d.n^{p+o(1)})$ query run time and \\$O(n^{1+p+o(1)} +d.n)$ for space, where $p=1/2c^2-1$ and $c>1$ in the Euclidean space and $p=1/2c-1$ for the Hamming space.

\begin{figure*}
\vspace{0.2cm}
\includegraphics[width=\textwidth,height=10cm]{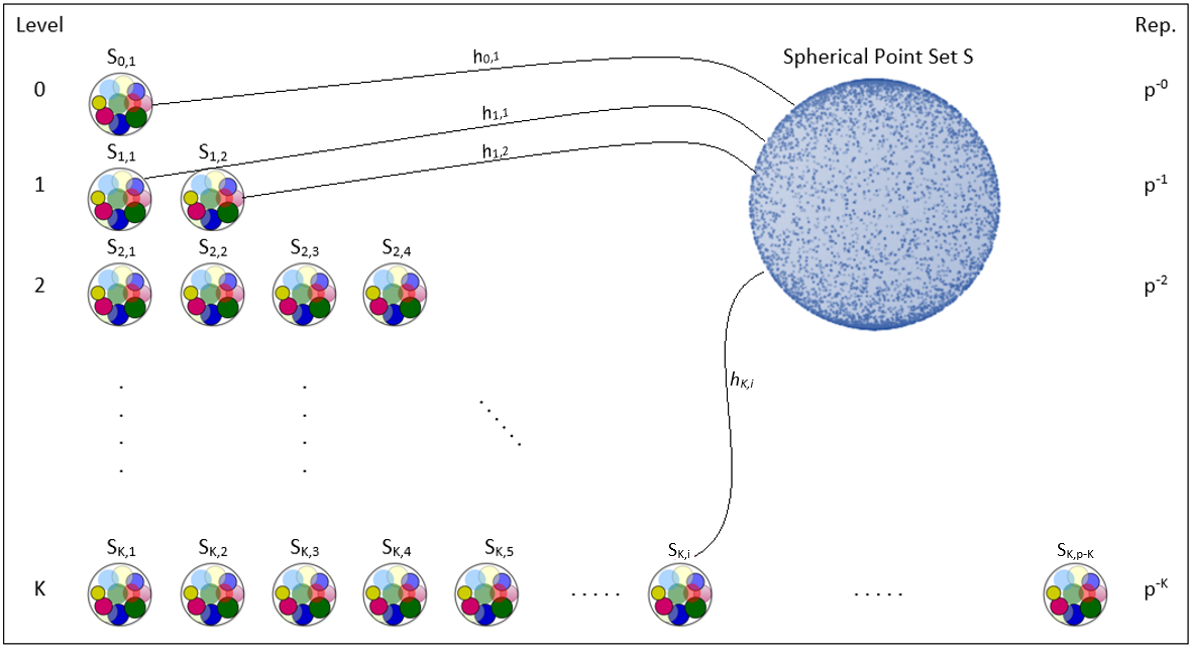}
\caption{ Overview of the spherical multi-level data structure}
\end{figure*}

\section{Data Structure}
In this section we discuss, to our knowledge the first parameter-free, multi-level data-dependant LSH data structure. Our data structure is a modified version of the data-dependant Spherical LSH data structure created by Andoni and Razenshteyn \cite{Andoni2015a} and uses the multi-level data structure approach as presented by Thomas et al\cite{Ahle2016}.

\subsection{Overview}

Our modified data structure is a 2-dimensional Spherical LSH structure. The building algorithm takes the following parameters, $D_{o}$ and $L$; where $D_{o}$ is a given data set, which we assume lies in the Euclidean space and contains n points and $L$ is an upper bound threshold which limits the number of $k$ repetitions used to build each level of the data structure. We further assume this space has a dimension of $d = \Theta(log n * log log n)$ \cite{Andoni2015a}. 

We made the following modifications to the "PROCESS" \cite{Andoni2015a} function in the Spherical LSH data structure building algorithm in order to make Spherical LSH a mutli-level data structure:
\begin{itemize}
  \item Calculate a parameter $k=\ceil*{log n / log(1/p2)}$, repetitions. Where $p_{1}$ and $p_{2}$ are the probabilities of collision for close points and probability of collisions for far points respectively.
  \item Calculate the number of repetitions needed $numreps = \ceil*{p_{1}^-k}$
  \item Initialize a 2-dimensional data structure $M$ to hold the $k$ layers of data points partitioned into spheres.
\end{itemize}

The initial partitioning of the data set $D_{o}$, partitions the data into subsets components, that are then enclosed into balls and eventually these balls are added to spheres $S_{K}$. We iterate from $i=0 \quad to \quad numreps$ over all the balls that have been created. We also keep a running count of the number of spheres $S_{K}$ that have been created. For each new level (repetition) we add $i + 1$ spheres $S_{K}$ to $M_{i}$ level.
 
\section{Querying Algorithm}
The multi-probing technique we employ is a common querying strategy for LSH data structures. The idea behind this technique is to make sure each bucket does not have the same or similar high collision probability as other buckets.

We leveraged Algorithm 1 which was presented by \cite{Ahle2016} in order to implement the multi-probing querying technique for our data structure.

\begin{algorithm*}[h]
\caption{Adaptive-Multi-probe(q,$\sigma$, S)}
\label{euclid}
\begin{algorithmic}[1]
\State $w_{best} \gets n;  k_{best} \gets 0; \jmath_{best} \gets 1$
\State $PQ \gets \textbf{empty priority queue} $ 
\newline $\quad\triangleright$ Manages pairs (k, $\jmath$) with priority cost(k, $\jmath$) = $\jmath$.reps(k, $\jmath$).
\State $PQ.insert((1, 1))$
\While {$PQ.min() < w_{best}$} 
\State $(k,\jmath) PQ.extractMin()$
\If{$k < K \quad and \quad \jmath = 1$}
\State $PQ.insert((k + 1, 1))$
\EndIf
\State $PQ.insert((k, \jmath + 1))$
\State $w_{k,\jmath }\gets  $$\sum_{i=1}^{reps(k,\jmath )}\sum_{j=1}^{\jmath } (1 + |S_{k,i,j}(q)|) $$  $
\If{$w_{k,\jmath} < w_{best}$}
\State $k_{best} \gets k; \jmath_{best} \gets \jmath; w_{best} \gets w_{k,\jmath} $
\EndIf
\EndWhile
\State \textbf{return} $ \bigcup_{i=1}^{reps(k_{best},\jmath_{best})} \bigcup_{i=1}^{\jmath_{best}} \{x \in S_{k_{best},i,j}(q) | dist(x, q) \leq r\} $
\newline \newline
Adaptive multi-probing query algorithm on the Multi-Level Spherica LSH data structure with K levels and reps(k,$\jmath$) $= \lceil 2 log(2\jmath k)/P_{k,\jmath}\rceil.$ For clarity, we write $S_{k,i,j}(q) = S_{k,i}(\sigma_{k,j}(h_{k,i}(q)))$ for the $j^{th}$ bucket in the sequence at level k, repetition i \cite{Ahle2016}.
\end{algorithmic}
\end{algorithm*}

\section{Future Work}
Our research are leveraging the theoretical guarantees of the papers which our work is based. More work is needed to formally prove that Multi-Level Spherical LSH data achieves $O(n^p + t)$ query run time, for all inputs n where $t <= n$. We will also implement a library to test the practical feasibility of our data structure.

\bibliographystyle{abbrv}
\bibliography{mendeley}  

\end{document}